# Nanoscale magnetic and charge anisotropies at manganite interfaces

Santiago J. Carreira,[a,b] Myriam H. Aguirre,[c,d,e] Javier Briatico,[f] and Laura B. Steren,[a,b]

Strong correlated manganites are still under intense research owing to their complex phase diagrams in terms of the Sr-doping and their sensitivity to intrinsic and extrinsic structural deformations. Here, we performed x-ray absorption spectroscopy measurements of manganites bilayers to explore the effects that a local Sr-doping gradient produce on the charge and antiferromagnetic anisotropies. In order to gradually tune the Sr-doping level along the axis perpendicular to the samples we have grown a series of bilayers with different thicknesses of low-doped manganites (from 0 nm to 6 nm) deposited over a $La_{0.7}Sr_{0.3}MnO_3$ metallic layer. This strategy permitted us to resolve with high accuracy the thickness region where the charge and spin anisotropies vary and the critical thickness $t_c$ over which the out of plane orbital asymmetry does not have any further modifications. We found that the antiferromagnetic spin axis points preferentially out of the sample plane regardless the capping layer thickness. However, it tilts partially into the sample plane far from this critical thickness, owing to the jointed contributions of the external structural strain and electron doping. Furthermore, we found that the doping level of the capping layer sensibly affects the critical thickness, giving clear evidence of the influence exerted by the electron doping on the orbital and magnetic configurations. These anisotropy changes induce subtle modifications on the domain reorientation of the $La_{0.7}Sr_{0.3}MnO_3$, as evidenced from the magnetic hysteresis cycles.

[a.] Consejo Nacional de Investigaciones Científicas y Técnicas, Argentina. Tel: 54-11 6772-7103; E-mail: *steren@tandar.cnea.gov.ar*.
[b.] Laboratorio de Nanoestructuras Magnéticas y Dispositivos. Dpto. Materia Condensada e Instituto de Nanociencia y Nanotecnología (INN), Centro Atómico Constituyentes (CNEA), (1650) San Martín, Buenos Aires, Argentina.
[c.] Instituto de Ciencia de Materiales de Aragón (ICMA) e Instituto de Nanociencia de Aragón (INA), Universidad de Zaragoza, E-50018 Zaragoza, Spain. Fax: +34 976 76 2776; Tel: +34 876 55 5365; E-mail: *maguirre@unizar.es*.
[d.] Departamento de Física de la Materia Condensada, Universidad de Zaragoza, E-50009 Zaragoza, Spain.
[e.] Laboratorio de Microscopías Avanzadas, Universidad de Zaragoza, E-50018 Zaragoza, Spain.
[f.] Unité Mixte de Physique, CNRS, Thales, Université Paris-Sud, Université Paris-Saclay, Palaiseau 91767, France.

## 1. Introduction

Artificial heterostructures of complex oxides are still highly attractive as an alternative path to overcome the miniaturization issues arisen in current semiconductor technologies. Among these oxides, perovskite manganites offer an excellent illustration of the rich and complex electronic and magnetic phases that can arise at oxides interfaces, due to the strong interrelation between the charge, spin and lattice degrees of freedom[1–4]. Owing to their colossal magnetoresistance and half metallicity[4,5], these materials are considered an excellent source of spin polarized currents, with potential applications in magnetic field sensors, magnetic memories and spintronic devices[6–8]. Moreover, the compound $La_{0.7}Sr_{0.3}MnO_3$ ($LS_{0.3}MO$) is usually employed as a spin injector in spin valves[9] and in magnetic tunnel junctions[10].

In epitaxial thin films of manganites, the symmetry of the Mn 3d orbitals and the magnetic exchange interactions are strongly affected by the strains induced by the substrate, stabilizing unique electronic phases.

As a general trend, a compressive strain induces an elongation of the oxygen octahedral along the axis perpendicular to the sample plane, stabilizing the occupation of the $e_g$ $3z^2-r^2$, whereas a tensile strain promotes a compression of the oxygen octahedral, favoring the $x^2-y^2$ filling[11,12]. In addition to the strain effects, the atomic plane termination at manganite free surfaces, namely $MnO_2$ or La/SrO, can locally alter the orbital filling. For instance, the absence of the apical oxygen in the $MnO_2$ plane termination may stabilize the $3z^2-r^2$ orbital occupation, even under a tensile strain environment[13]. The contribution of the spin density distribution to the surface magnetic anisotropy plays also a decisive role in the resulting orbital configuration, in addition to the strain induced by the substrate[14]. The scenario becomes even more rich and complex when we consider Sr-doping gradients with characteristic lengths ranging in the nanometer scale. Such situation deserves particular attention in thin films of lightly Sr-doped manganites, as they exhibit, according to the bulk phase diagram, magnetic and transport phase transitions within the doping range $0 \leq x \leq 0.1$. The high sensitivity to slightly doping modifications becomes evident in thin films of $La_{0.9}Sr_{0.1}MnO_3$ ($LS_{0.1}MO$), where their magnetism and transport properties are enhanced with respect to their bulk counterpart. Moreover, they exhibit a thickness dependence of the metal to insulator transition driven by strain relaxation and off-stoichiometry effects[15].

We recently reported the evolution of the ratio $Mn^{3+}/Mn^{4+}$ in terms of the barrier thickness and found that the $Mn^{3+}$ content increases as a consequence of the larger contribution of the low-doped manganite barrier to the signal[16]. The magnetic order in manganite thin films is not purely ferromagnetic, but presents antiferromagnetic phases coexisting with the ferromagnetism and the contribution of each magnetic order is governed mainly by the strain and the bandwidth of the Mn 3d orbitals[17,18]. This feature becomes particularly important at the interface between manganites with different Sr-dopings, because the bandwidth and transport properties are severely affected by the $Mn^{3+}/Mn^{4+}$ ratio[19]. This scenario motivated us to explore in detail the nature of the local antiferromagnetic anisotropy in manganite thin films with Sr-doping gradients, where orbital and charge reconstructions take place simultaneously within a few units cells.

We are particularly intrigued in studying the local magnetic anisotropy at manganites' surfaces, where a Sr-doping gradient is artificially altered along the axis perpendicular to the sample plane, creating a spatial doping gradient which can locally modify their interfacial electronic properties. To this goal, we explored the orbital and magnetic phases throughout manganite interfaces $La_{0.7}Sr_{0.3}MnO_3/La_{1-x}Sr_xMnO_3$ ($LS_{0.3}MO/LS_xMO$) with x = 0 and x = 0.1, in order to correlate the local changes in the composition with the structural distortions, orbital ordering and magnetic anisotropy. As a surface-sensitive technique, x-ray absorption spectroscopy has proven to be the most adequate tool to reveal both orbital occupancy and the existence of complex magnetic phases[11,12,20–23]. In order to describe the $LS_{0.3}MO/LS_xMO$ interfaces and determine the evolution of their properties as the $LS_xMO$ capping layer thickness increases, we studied bilayers with different $LS_xMO$ thicknesses, in the range 0 nm ≤ t ≤ 6 nm (hereafter refer as t-$LS_xMO$ series, where t is the thickness and x is the nominal Sr-doping of the capping layer). These experiments allowed us to establish characteristic length scales for the spatial evolution of the anisotropy in the orbital filling of the 3d energy levels. Remarkably, the spatial evolution of the antiferromagnetic axis presents a non-trivial behaviour, with the highest out of plane anisotropy at intermediate Sr-doping gradients.

These intriguing properties promote subtle effects on the overall magnetic anisotropy of the films, which are revealed by the coercive field dependence observed in terms of the capping layer thickness t. The competition between the strain and charge doping at the surface gives rise to a complex spatial profile of the overall magnetic anisotropy profiles.

These findings provide an outstanding progress in the field of nanoscale magnetism, as they determine quantitatively length scales for the orbital and magnetic reconstruction at manganite surfaces. The direct link of these features and the evolution of the lattice deformation and local stoichiometry are unambiguously verified, as well as their link with the overall magnetic domain reorientation of the $LS_{0.3}MO$.

## 2. Results and discussion

### 2.1 Structural and chemical analysis

The symmetry of the crystalline structure in manganites is severely affected by the Sr-doping level and subtle structural distortions of the oxygen octahedral strongly alter the resulting electronic and magnetic properties. In this work we use as the inner layer the half-metallic manganite $LS_{0.3}MO$, which in bulk crystallize in a rhombohedral structure $R\bar{3}c$ with a pseudocubic lattice parameter $a$ ($LS_{0.3}MO$) = 3.875 Å[24]. The outer layers, however, consist of low-doped manganites $LS_xMO$ which in bulk exhibit an orthorhombic structure. The pseudocubic lattice constant for the un-doped compound $LaMnO_3$ is $a$ (LMO) = 3.94 Å[25], whereas for the low-doped counterpart $LS_{0.1}MO$ is $a$ ($LS_{0.1}MO$) = 3.92 Å[26]. When the $LS_{0.3}MO$ and $LS_xMO$ are brought together in a thin film, the differences in the unit cell size will produce local strain modifications in the near surface region and these modifications can be revealed with a strain analysis of HRSTEM-HAADF images.

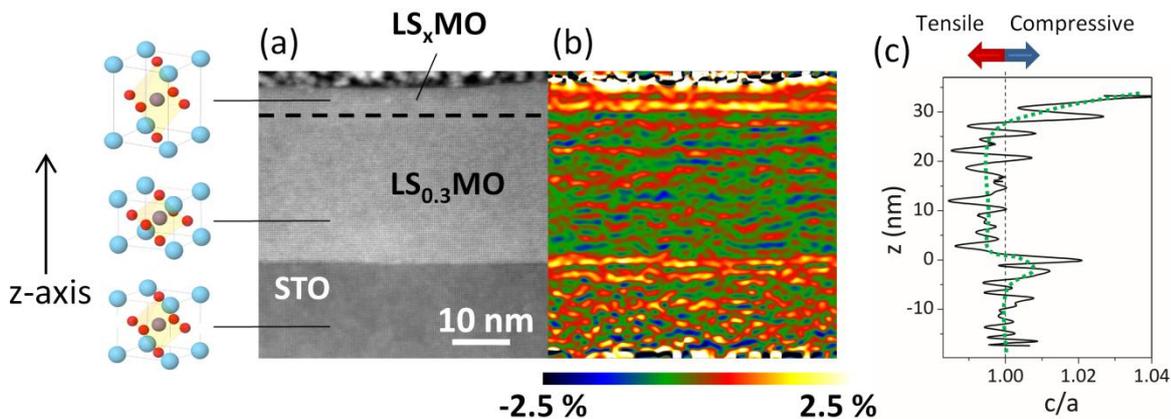

**Fig. 1:** (a) Cross sectional STEM-HAADF image of a STO//$LS_{0.3}MO$/$LS_xMO$ bilayer. The $LS_{0.3}MO$/$LS_xMO$ interface is indicated by a dashed line. (b) GPA strain analysis map of the deformation perpendicular to the film plane. (c) Estimated c/a ratio along the z-axis, extracted from the analysis of the GPA image. A vertical dashed line separates the Tensile (c < a) from the compressive (c > a) regions.

In Fig. 1(a) we show a typical STEM-HAADF image of a bilayer, where the z-contrast at the interface with the $SrTiO_3$ substrate is clearly visible and no interdiffusion effects were detected. On the contrary, the chemical contrast at the interface of $LS_{0.3}MO$/$LS_xMO$ is too small to be visible in the HRSTEM images. The presence of the low-doped manganite at the outer layers is revealed by assessing the structural deformation along the perpendicular direction with Geometrical Phase Analysis (GPA) on the STEM images[27]. In the strain map shown in Fig. 1(b), three distinct regions can be observed along the z-axis, corresponding to different out of plane lattice parameters c. Taking into account that the in-plane parameter a of both layers matches with those of the substrate, we found that the $LS_{0.3}MO$ unit cell is compressed with respect to the $SrTiO_3$ substrate (c/a < 1) and, on the contrary, the LS$_x$MO is elongated along the z-axis. Therefore, a clear transition from a compressed oxygen octahedral with $c/a < 1$ to an elongated oxygen octahedral with $c/a > 1$ is present at the interfaces LS$_{0.3}$MO/LS$_x$MO. Based on the GPA images and considering coherent in plane deformations we can deduce the mean unit cell deformation $c/a$ along the z-axis, relative to the substrate, which is plotted in Fig. 1(c). This analysis were complemented with standard x-ray diffraction measurements (further details in Fig. S1 of Supp. Info.), in order to calculated the out of plane lattice constant of the LS$_{0.3}$MO for both series and to have an estimation of the unit cell distortion of both layers. The structural data is resumed in Table 1, where we also included the vertical deformation of both layers relative to their bulk counterparts $\varepsilon_{ZZ} = 100\,(c - c_{bulk})/c_{bulk}$. For both series we found that the LS$_{0.3}$MO is contracted and the LS$_x$MO is expanded with respect to their bulk values, which further confirms the presence a local strain gradient with important implications on the subsequent magnetic behavior.

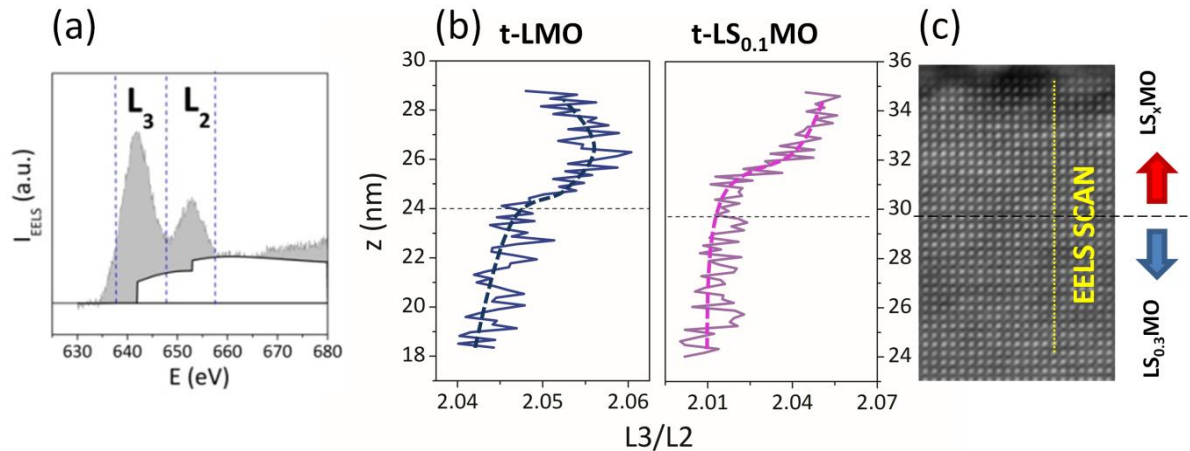

**Fig. 2:** (a) EELS spectra at the Mn L edge. (b) L$_3$/L$_2$ intensity ratio extracted from the EELS scans throughout the interfaces LS$_{0.3}$MO/LS$_x$MO, along the direction specified in (c). The horizontal dash lines in (b) and (c) indicate the interface between the LS$_{0.3}$MO inner layers and the LS$_x$MO capping layers.

| Serie | LS$_{0.3}$MO | | | LS$_x$MO | | |
|---|---|---|---|---|---|---|
| | c (Å) (±0.05) | $\varepsilon_{zz}$ (%) | c/a | c (Å) (±0.05) | $\varepsilon_{zz}$ (%) | c/a |
| t - LMO | 3.87 | -0.1 | 0.991 | 3.95 | 0,3 | 1.012 |
| t - LS$_{0.1}$MO | 3.86 | -0.4 | 0.988 | 3.94 | 1 | 1.010 |

**Table 1** Structural parameters obtained for the bilayers. Out of plane lattice parameters c, vertical deformation $\varepsilon_{zz}$ and c/a ratio for the LS$_{0.3}$MO and LS$_x$MO layers of both series

From the electronic perspective, the presence of the low-doped manganites at the surface will supply electrons to fill the 3d shells of the Mn ions, altering the Mn$^{3+}$/Mn$^{4+}$ ratio. Atomic resolution Electron Energy Loss Spectroscopy (EELS) analysis of the STEM images is a valuable tool to probe, with spatial resolution, the oxidation state of the Mn in the near surface region. The L edges of the Mn ions results from excitations of 2p electrons into 3d empty bound states or the continuum. Due to the spin-orbit splitting of the 2p state, the spectra shows two absorption peaks L$_3$ and L$_2$, which can be found around 644 eV and 655 eV respectively, and are related to the transitions $2p_{3/2} \rightarrow 3d_{5/2}$ and $2p_{1/2} \rightarrow 3d_{3/2}$, as depicted in Fig. 2(a)[28]. Experimentally, it is found that the intensity ratio between these lines L$_3$/L$_2$ increases with the number of electrons in the 3d bands[29]. In Fig.2(b,c) we show the ratio L$_3$/L$_2$ for one sample of each series along a vertical line scan, which reveals that the oxidation state of the Mn is reduced as we approach to the surface. As was expected, the presence of the low-doped manganites at the outer layers increase the Mn$^{3+}$/Mn$^{4+}$ ratio, reducing the overall Mn valence.

## 2.2 X-ray linear dichroism experiments

X-ray absorption spectroscopy (XAS) experiments in total electron yield mode (TEY) at the Mn L-edge were performed to disentangle the contributions of different orbital and magnetic states at the interfaces, where Sr-doping gradients are present. The experimental geometry used in shown in Fig. 3. The incidence angle of the photon beam in XAS measurements was 30° with respect to the sample surface. The electric field of the incident linear polarized radiation lies into the sample plane for the horizontal polarized radiation (HPR) and almost perpendicular to the sample plane for the vertical polarized radiation (VPR).

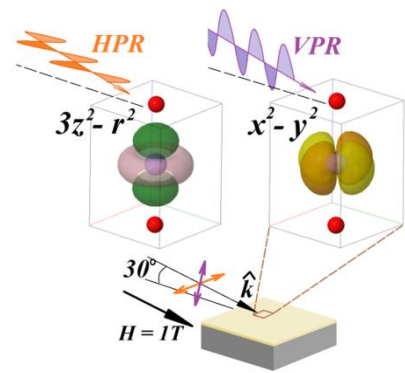

**Fig. 3:** Scheme of the experimental geometry. The orientation of the Mn $e_g$ orbitals $x^2-y^2$ and $3z^2-r^2$ with respect to the polarized radiation are indicated.

To determine the orbital occupancy and the magnetic structure at the interfaces we performed linearly polarized x-ray absorption spectroscopy (XLD). The linearly-polarized x-ray absorption detects the charge anisotropy along the direction of the electric field, measuring the number of valence holes. In magnetic materials, the magnetization axis is detected through the spin-orbit coupling of the spins to the lattice. The XLD spectrum includes, therefore, orbital (XNLD) and magnetic (XMLD) contributions[22,30]. To disentangle the orbital from magnetic dichroisms in magnetic materials, XAS experiments are performed at different temperatures. The XNLD is assumed to be temperature independent and so it will be the only effect that could be causing a dichroic signal close and above the magnetic ordering temperature[27].

In Fig. 4 (a,b) we present typical XLD spectra for t-LMO and t-LS$_{0.1}$MO samples measured at three different temperatures:

$$T_1(T \ll T_c^{LS_xMO}) < T_2(T \sim T_c^{LS_xMO}) < T_3(T \sim T_c^{LS_{0.3}MO})$$

where $T_c^{LS_{0.3}MO}$ and $T_c^{LS_xMO}$ are the ordering temperature of the LS$_{0.3}$MO and capping layers respectively. $T_1$ and $T_3$ have been kept at 4 K and 300 K respectively, while $T_2$ was set to 150 K and 230 K for t-LMO and t-LS$_{0.1}$MO series of samples.

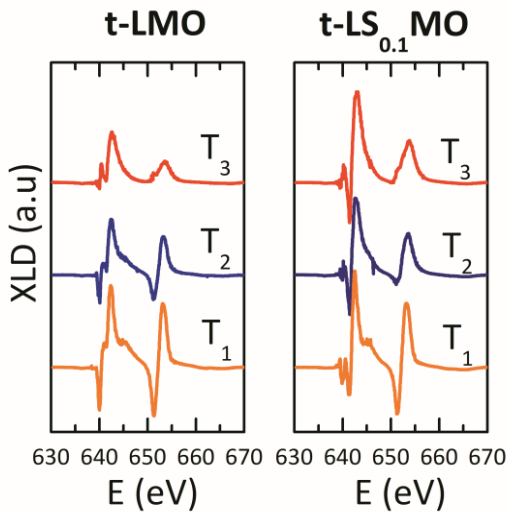

**Fig. 4:** Typical XLD spectra for one sample from the t-LS$_{0.1}$MO (solid lines) and t-LMO (dash lines) sample series. Spectra measured at low temperatures far below $T_c^{LS_xMO}$, $T_1$; around $T_c^{LS_xMO}$, $T_2$; and around $T_c^{LS_{0.3}MO}$, $T_3$. The curves at different temperatures were shifted vertically.

### 2.2.1 Preferential orbital occupancy

In order to probe the orbital occupation asymmetry, we evaluated the XAS spectra obtained at room temperature. As is shown in Fig. 3, the electric field was polarized nearly parallel to the [001] crystallographic axis (VPR) and perpendicular to it (HPR). The XLD was calculated as the difference between XAS spectra measured with HPR and VPR polarization. According to the XNLD definition and considering the symmetry of the $e_g$ orbitals $3z^2$-$r^2$ and $x^2$-$y^2$, the dichroism will be positive (negative) for a preferential orbital occupation $3z^2$-$r^2$ ($x^2$-$y^2$). In this regard, we found that the XNLD spectra are positive over almost the entire energy range for both series of samples, suggesting that the number of occupied states at the $3z^2$-$r^2$ orbitals is larger than those of the $x^2$-$y^2$ orbitals.

The area under the XNLD spectra is proportional to the charge anisotropy and its sign defines the electronic preferential occupation asymmetry. We calculated the integral of the XNLD spectra in the energy range between 630 eV and 670 eV, $I_{XNLD\,Mn}$, to explore the orbital occupation asymmetry across the interfaces. The $I_{XNLD\,Mn}$ of the XNLD bilayers spectra is plotted as a function of the barrier thickness in Fig. 5. As a general trend, this intensity increases with the barrier thickness for both set of samples, consistent to an increasing electronic population at the $3z^2$-$r^2$ orbitals as the barrier thickness increases.

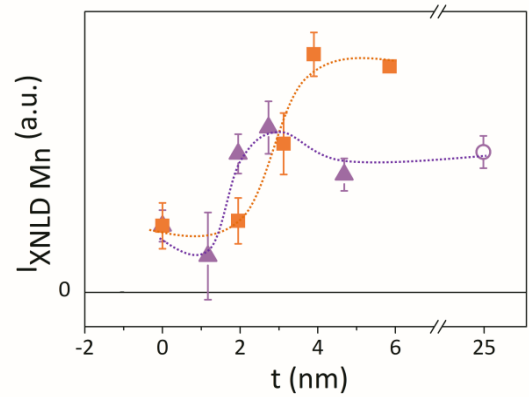

**Fig. 5:** Thickness dependence of the $I_{Mn\,XNLD}$ for the (triangles) t-LMO and for the (squares) t-LS$_{0.1}$MO sample series. The value corresponding to the LaMnO$_3$ reference sample (open circle) is also shown. Dashed lines serve as a guide to the eye.

It is noteworthy that $I_{XNLD\,Mn}$ is positive even for the LS$_{0.3}$MO reference sample, where their surface structure is still subject to a tensile strain ($c < a$). This could be originated by the absence of the apical oxygen characteristic of the MnO$_2$ surface termination, as revealed by the RHEED patterns measured in-situ after growth (see Fig. S3 of the Supp. Info.). For the thicker barriers the $I_{XNLD\,Mn}$ reaches a plateau, indicating that the orbital asymmetry does not increase any further. Moreover, the $I_{XNLD\,Mn}$ of the LaMnO$_3$ surface shows similar values to those obtained with the thicker barriers of the t-LMO series, suggesting that the orbital configuration of the LaMnO$_3$ surface resemblance to those observed for the bilayers with the thicker capping layers. These results show that the $3z^2$-$r^2$ preferential orbital occupation at the outer layers increases gradually within a thickness range that depends on the doping, being ~ 3 nm for the t-LMO and ~ 4 nm for the t-LS$_{0.1}$MO, and saturates beyond a critical thickness $t_c$ of the low-doped capping layer. This charge anisotropy is also reflected in the XNLD spectra obtained at the O-K edge, which is sensitive to dipolar transitions from the O-1s to O-2p energy levels (Fig. 6)[32,33,34].

The XLD spectra at this edge keeps always positive values for the rest of the bilayers studied, regardless of the capping layer thickness, signaling that the orbital 2p$_z$ of the oxygen ion, whose lobes are parallel to the [001] axis, are strongly hybridized to the $3z^2$-$r^2$ orbitals promoting the higher orbital occupation and electron delocalization along the [001] direction. As is sketched at the inset of Fig. 6, both Mn L-edge as well as O-K edge unveils a clear charge anisotropy in the orbital filling.

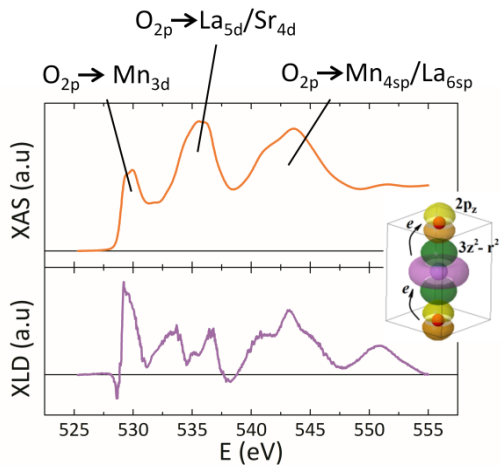

**Fig. 6:** Typical XAS (upper panel) and XLD (lower panel) spectra at the O-K edge for one bilayer of the t-LMO series. The three absorption regions related to the hybridizations $O_{2p} \rightarrow Mn_{3d}$, $O_{2p} \rightarrow La_{5d}/Sr_{4d}$ and $O_{2p} \rightarrow Mn_{4sp}/La_{6sp}$ are indicated. The cartoon schematizes the orbital hybridization between the $2p_z$ of the apical oxygen ions and the Mn $3z^2-r^2$ orbitals.

### 2.2.1 Antiferromagnetic anisotropy and magnetic pinning effects

The magnetic contribution to the XLD, both ferromagnetic and antiferromagnetic, emerges well below the ordering temperature. In order to discern between these contributions and analyze the antiferromagnetic phase individually, a magnetic field of 1 T was applied to the samples in the direction of the incident x-ray in order to saturate and cancel out the ferromagnetic contribution from the dichroic signal (further details in Fig. S3 of the Supp. Info.). The XMLD was then calculated as the difference of the low temperatures and the room temperature XLD spectra[35]. The resulting XMLD spectrum is mainly negative, meaning that the absorption with VPR is larger than the absorption with HPR, in agreement with an antiferromagnetic spin direction oriented along the [001] axis.

In order to assess the spatial evolution of the antiferromagnetic phases when the barrier thickness increases we calculated the area under the XMLD spectra, $I_{AF}$, as a measure of the antiferromagnetic anisotropy along the [001] axis. In Fig. 7 we plotted the thickness dependence of the $I_{AF}(T)$ calculated between 649.7 eV and 652.7 eV for both sample's series and for $T_1$ and $T_2$. The calculation of $I_{AF}(T)$ in a wider energy range does not change the general trend, probing the robustness of these experimental results.

Note that $I_{AF}$ present a minimum value at intermediate capping layer thicknesses and takes values closer to zero (although still lower than zero) away from this intermediate thickness range. In terms of the experimental geometry considered here we deduced that the antiferromagnetic spin axis has a strong out of plane (OOP) component at this intermediate thickness, that partially tilts into the sample plane for capping layers thickness different from this critical value.

We should also keep in mind, however, that $I_{AF}$ is mainly negative, meaning that the spin alignment always keeps an OOP component that prevails. Recalling the evolution of the preferential orbital occupation discussed in Fig. 5 and Fig. 6, it is important to emphasize that the critical thickness $t_c$ where the orbital anisotropy saturates matches with the thickness value where $I_{AF}$ is minimum. Clearly the evolution of the orbital and antiferromagnetic contributions across the $LS_{0.3}MO/LS_xMO$ interfaces are intimately related and should be analyzed consistently, based on a single model.

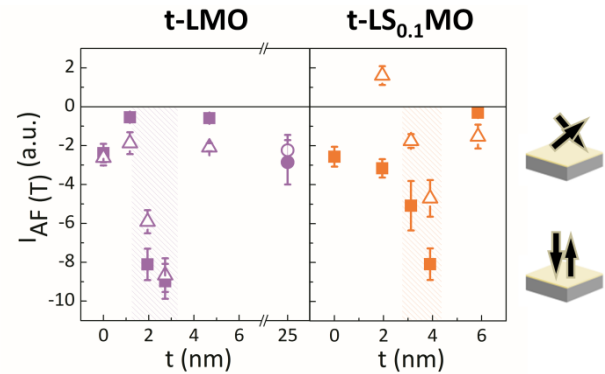

**Fig. 7:** Thickness dependence of $I_{AF}$ for the t-LMO and t-LS$_{0.1}$MO series, respectively. Measurements at 4 K (filled squares) and at 150 K (t-LMO) / 230 K (t-LS$_{0.1}$MO) (open triangles) are shown. On the right side of the graph, we represented with schemes the orientation of the antiferromagnetic spins for each value of $I_{AF}$.

As a first approach, the evolution of the orbital and antiferromagnetic anisotropies can be correlated with the valence changes across the interface. The extra electrons supplied by the low-doped manganites modifiy gradually the $Mn^{3+}/Mn^{4+}$ ratio and fill preferentially the out of plane $e_g$ orbitals lying at lower energies, due to the broken symmetry induced by the compressive strain. The charge anisotropy in the [001] direction joint with the tetragonal distortion induced by the substrate reinforce the orbital hybridization and the ferromagnetic interaction along this axis and simultaneously the antiferromagnetic coupling within the (001) plane. As a result, the antiferromagnetic axis tilts locally out of sample plane and the charge carriers are delocalized with a preferred orientation.

The existence of the critical thickness, however, may be interpreted using different approaches. First, we need to consider that the probing region of the x-ray beam in TEY mode (~ 4 nm) is similar to the capping layer thickness[36]. Therefore, we can assume that above this critical thickness the XAS signal is no longer sensitive to the interface $LS_{0.3}MO/LS_xMO$ but probes mainly the barrier properties. This argument is consistent with the overall Mn valence profile, which does not show any further modifications once reached the critical barrier thickness $t_c$[16]. The out-of-plane orbitals are fully occupied above this critical thickness and the extra electrons begin to fill the in-plane $x^2-y^2$ orbitals, which partially disrupts with the OOP anisotropy of the orbital filling and OOP antiferromagnetic spin alignment. The local oxygen octahedral rotations characteristic of the lightly doped manganites[37] are expected to increase at the outer layers and may compete with the coherent distortions of the unit cell induced by the substrate, favoring distinct electronic configurations.

Additionally, we found that the orbital and antiferromagnetic anisotropies for the thicker barriers of the t-LMO series are similar to those measured for the LaMnO$_3$ surface. Furthermore, the critical thickness of the t-LMO series (~ 3 nm) is lower than the one observed for the t-LS$_{0.1}$MO series (~ 4 nm). This is an expectable result considering that the proportion of $e_g$ electrons supplied by the undoped LaMnO$_3$ is larger than those supplied by the low doped La$_{0.9}$Sr$_{0.1}$MnO$_3$. Consequently, the critical thickness needed to saturate the OOP orbital occupation should be lower for the t-LMO sample'series.

The local structural distortion of the oxygen octahedral evolves from a tensile strain at the inner layers to a compressively unit cell at the outer layers. However, the anisotropy of the antiferromagnetic spin axis as well as the preferential orbital filling does not show the same trend. This observation gives clear evidence that in systems with a local variation of the electron doping joint with local structural distortions, the final orbital and magnetic configurations results from the interplay between external factors like the strain state and atomic surface termination as well as intrinsic features like the $Mn^{3+}/Mn^{4+}$ ratio driven by the Sr-doping.

We could also envisage a complementary approach to explain these findings. Following the arguments discussed by D. Pesquera et al.[17] and the complex phase diagram of the bulk Sr-doped manganite compounds, the saturation of the preferential orbital occupation above a critical thickness may also be interpreted in terms of local bandwidth changes across the interface. In contrast to the large bandwidth and metallic properties exhibited by the $LS_{0.3}MO$ at the inner layers, the lightly or undoped $LS_xMO$ capping layers present narrower bands with electrons more localized. In materials with a reduced electron mobility the effects of strain on the orbital anisotropy and antiferromagnetic phases tends to be reduced as well, which is revealing in the unmodified XNLD and XMLD integrals for the sample with the thicker capping layers.

Complementary to the local magnetic properties, it is important to explore whether these local anisotropy changes at the surface may be reflected on the overall magnetic properties of the films, when magnetic hysteresis cycles are performed. To this end, we measured hysteresis cycles at 4 K with an MPMS-SQUID of both sample'series with the magnetic field oriented in-plane and compared the magnetization reversal process (further details on the magnetic cycles can be found in Fig. S4-S6 of the Supp. Info.).

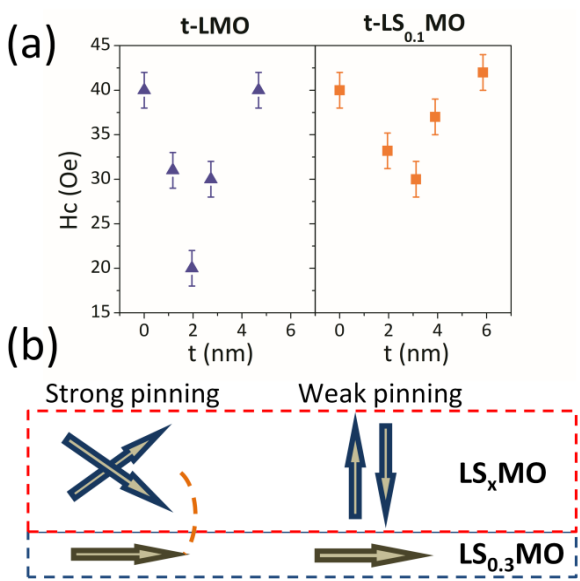

**Fig. 8:** (a) In-plane coercive field vs thickness for t-LMO and t-$LS_{0.1}$MO series. (b) Sketch of the physical model based on the magnetic pinning effect.

In Fig. 8(a) we plotted the evolution of the coercive field in terms of the capping layers thickness, where it becomes evident that subtle changes in the magnetization reversal arise and the minimum coercive field agrees with the critical thickness discussed before.

This result may appear counterintuitive at first because lower in-plane coercive fields are found when larger out of plane antiferromagnetic anisotropy takes place, but a simple model based on magnetic pinning of domains can explain these phenomena. As sketched in Fig. 8(b), we argue that the in-plane component of the antiferromagnetic order at the surface turns out to be more effective to pin the neighboring $LS_{0.3}MO$ magnetic domains and prevent the reversal of the $LS_{0.3}MO$ magnetization under the application of an external magnetic field. The effects on the overall magnetic behavior are subtle and will depend on the nature and coherence of the antiferromagnetic order over large areas. Still, this findings provide a exciting platform to correlated local and macroscopic properties in manganite thin films.

## 3. Conclusions

In summary, we found a direct correlation between local antiferromagnetic spin axis and the preferential orbital occupation throughout the interfaces $LS0.3MO/LS_xMO$. The antiferromagnetic anisotropy reaches a maximum value with the spin axis oriented OOP for intermediate barrier thicknesses, giving rise to a strong ferromagnetic interaction along this axis and a large orbital hybridization mediated by the eg ($3z^2-r^2$) levels. In addition, we quantified the barrier thickness'range where the electronic reconstruction takes place and found that apart from this critical thickness $t_c$ the antiferromagnetic spin axis tilts partially away from the [001] axis. This critical thickness arises from the contribution of the local strain, which stabilize the OOP electronic occupation for an axial expansion of the unit cell and Sr-doping of the capping layers, which determine the amount of extra electrons supplied at the interface. The spatial variation of the antiferromanetic anisotropy brings subtle change on the overall magnetization reversal process of the $LS0.3MO$, as revealed from hysteresis cycles.

These results provide valuable progresses in the field of strongly correlated oxide interfaces, where the structural, orbital and magnetic degrees of freedom can be tuned locally within a few unit cells.

## 3. Experimental section

Two series of $La_{0.7}Sr_{0.3}MnO_3/La_{1-x}Sr_xMnO_3$ (x = 0, 0.1) bilayers (refer as $LS_{0.3}MO/LS_xMO$) were grown on (001) $SrTiO_3$ single-crystalline substrates by pulsed laser deposition. The thickness of the $LS_{0.3}MO$ electrode was kept fix at around 22 nm in both samples'series while the capping layer thickness was varied between 1.2 nm ≤ t ≤ 4.7 nm in the $LS_{0.3}MO/LaMnO_3$ series (t-LMO) and 2 nm ≤ t ≤ 6 nm in the $LS_{0.3}MO/La_{0.9}Sr_{0.1}MnO_3$ (t-$LS_{0.1}$MO) one. In addition, 20 nm-thick $LS_{0.3}MO$ and 25 nm-thick $LaMnO_3$ thin films were grown in similar conditions to use as references. The films thickness and texture were checked by x-ray reflectivity, x-ray diffraction and the overall magnetic properties were checked with a 7 T-SQUID magnetometer. The structural quality of the samples was assessed locally with scanning transmission electron microscopy coupled with a high angle annular dark field detector (STEM-HAADF) in a FEI Titan G2 at 300 keV probe corrected (a CESCOR Cs-probe corrector from CEOS Company) and fitted with a Gatan Energy Filter Tridiem 866 ERS to perform EELS analysis. Element-specific soft x-ray magnetic spectroscopy in total electron yield detection mode was used to probe the magnetic order and orbital occupation throughout the interface between the $LS_{0.3}MO$ electrode and the capping layers. The synchrotron radiation experiments were performed at the electron storage ring of the Helmholtz-Zentrum Berlin (BESSY II) by using the 7 T high-field end station located at the UE46-PGM1 beamline. We performed x-ray absorption spectroscopy (XAS) experiments at the L edge of Mn (630 eV to 670 eV) and at the O K-edge (525 eV to 555 eV).

## Conflicts of interest

There are no conflicts to declare.

## Acknowledgements

Authors thanks the financial support of FONCYT PICT 2014-1047, CONICET PIP 112-201501-00213, MINCYT and the European Commission through the Horizon H2020 funding by H2020-MSCA-RISE-2016 - Project N° 734187 - SPICOLOST. The microscopy studies have been conducted in the "Laboratorio de Microscopias Avanzadas" at "Instituto de Nanociencia de Aragon" - Universidad de Zaragoza and we thanks the access of equipments. Authors would like to acknowledge the use of Servicio General de Apoyo a la Investigación-SAI, Universidad de Zaragoza". Authors also acknowledge Dr. Eugen Weschke for fruitful discussions and the possibility to access to the Synchrotron facilitiy at BESSY II.

## Notes and references